\documentclass[a4paper,aps,pra,10pt,onecolumn,nofootinbib]{revtex4-2}
\RequirePackage[colorlinks,hyperindex]{hyperref}
\RequirePackage[english]{babel}
\RequirePackage[latin1]{inputenc}
\RequirePackage[T1]{fontenc}
\RequirePackage{mathrsfs}
\RequirePackage{amsmath}
\RequirePackage{amssymb}
\RequirePackage{amsbsy}
\RequirePackage{color}
\RequirePackage{bm}
\usepackage{bbm}
\usepackage{booktabs}
\hypersetup{colorlinks=true,breaklinks=true,urlcolor=blue,linkcolor=red}
\begin{document}
\title{\bf{A Covariant Framework for Generalized Spinor Dual Structures}}
\author{Rodolfo Jos\'e Bueno Rogerio,$^{\nabla}$\email{rodolforogerio@gmail.com}
Rogerio Teixeira Cavalcanti,$^{\gamma}$\!\! $^{\delta}$\email{rogerio.cavalcanti@ime.uerj.br}
Luca Fabbri,$^{c}$\!\! $^{\hbar}$\!\! $^{G}$\email{luca.fabbri@unige.it}}
\affiliation{$^{\nabla}$Centro Universit\'ario UNIFAAT, Atibaia-SP, 12954-070, BRAZIL\\
$^{\gamma}$Instituto de Matem\'atica e Estat\'istica, Universidade do Estado do Rio de Janeiro (UERJ), Rio de Janeiro-RJ,  20550-900, BRAZIL\\
$^{\delta}$Faculdade de Engenharia de Guaratinguet\'{a}, Universidade Estadual de S\~ao Paulo (Unesp), Guaratinguet\'{a}-SP,  12516-410, BRAZIL\\
$^{c}$DIME, Universit\`{a} di Genova, Via all'Opera Pia 15, 16145 Genova, ITALY\\
$^{\hbar}$INFN, Sezione di Genova, Via Dodecaneso 33, 16146 Genova, ITALY\\
$^{G}$GNFM, Istituto Nazionale di Alta Matematica, P.le Aldo Moro 5, 00185 Roma, ITALY}
\date{\today}
\begin{abstract}
In this work, we propose a novel framework for defining the dual structure of a spinor. This construction relies on the basis elements of the Clifford algebra, leading to a covariant structure that embeds the dual. The formulation includes free parameters that may be adjusted to meet specific requirements. Remarkably, it enables the explicit construction of representatives for each class within a recently proposed general classification of spinors. In addition to recovering known results, the formalism paves the way for the development of potential new theories in a manifestly covariant setting.
\end{abstract}
\maketitle
\section{Introduction}
The Dirac framework for spin-$1/2$ particles stands as one of the landmark accomplishments of contemporary physics. Nevertheless, despite its remarkable success and extensive scrutiny over the past century, a pivotal aspect of Dirac's formulation has remained relatively under explored until now: the dual structure intrinsic to spinors. The predictions of the theory are fundamentally tied to the conventional choice of this dual, consequently overlooking a broad spectrum of alternative formulations that might be instrumental in describing new fields beyond the Standard Model. The study of Elko spinors \cite{Ahluwalia:2004ab} directly addresses this oversight, drawing the community's attention to the potential of constructing, from fundamental principles, a viable framework for describing dark matter. Pursuing this direction has required imposing consistency conditions that, in turn, have motivated a rigorous and detailed mathematical examination of foundational aspects of Quantum Field Theory, with the aim of defining physical fields that encapsulate essential physical information.

The physical content of spinors is deeply tied to a fundamental aspect that is often overlooked: individual spinors are not directly observable. Instead, it is through their bilinear covariants composite quantities constructed from the spinor and its dual that physical observables emerge. The standard approach begins by defining the so-called Dirac dual of the spinor, followed by the construction of the corresponding bilinear covariants. This conventional framework for understanding spinor fields is widely presented in quantum field theory textbooks. Lounesto proposed a classification \cite{Lounesto:2001zz} scheme that organizes spinors into six mutually exclusive classes: including Dirac spinors (classes 1, 2, and 3), flag-dipole spinors (class 4), Majorana or flag-pole spinors (class 5), and Weyl or dipole spinors (class 6). The structure and relations among the bilinear covariants are governed by the Fierz-Pauli-Kofink (FPK) identities, which impose quadratic constraints that these quantities must satisfy. Crucially, the restriction to six distinct classes arises from Lounesto's use of the Dirac dual structure in his classification.

As we know, a significant part of the physical information is contained in the dual spinor. It is well known that the Dirac dual structure is not unique, nor is it the most fundamental, as is commonly presented in most textbooks. What happens is that the way it is introduced allows it to take on a compact form, usually denoted as $\bar{\psi} = \psi^{\dagger}\gamma_0$. As recently shown by Speran\c{c}a, in \cite{sperancca}, it is possible to define an operator that acts on the spinor, given by the  mathematical identity, namely $\mathcal{P}=m^{-1}\gamma_{\mu}p^{\mu}$, meaning that $\mathcal{P}\psi = \pm \psi$. Thus, when defining the dual structure, Dirac considered the following $\bar{\psi}=\psi^{\dag}\gamma_0$, responsible for establishing a Lorentz invariant quantity $\bar{\psi}\psi$. This justifies the (naive) omission of the operator $\mathcal{P}$ in textbooks.

However, as can be seen in the vast current literature on spinors, this form leads to numerous complications when the spinor in question is not a Dirac spinor for example, Elko spinors \cite{Ahluwalia:2004ab,Ahluwalia:2004sz}. In the specific case of Elko spinors, by employing a not well-defined dual structure for such spinors, one might lose information about the theory. As some problematic features, we can say that the Hamiltonian operator is Hermitian but possesses eigenstates with negative energy, the one-particle states associated with the quantum field expansion are not Hamiltonian eigenstates, violating the standard particle interpretation of QFT, besides locality is not reached \cite{gracia2025unraveling}.

In this way, it is common to consider the possibility of a reformulation of the dual structure, or some kind of freedom inherent in such a mathematical structure, in order to ensure that the physical information carried by the theory is preserved, and also to guarantee that the bilinear structures retain their character under specific Lorentz group transformations. These facts have been thoroughly scrutinized in the references \cite{Rogerio:2024lfh,HoffdaSilva:2019ykt,CoronadoVillalobos:2020yvr,Rogerio:2023kcp, gracia2025unraveling,BuenoRogerio:2017lim}

Thus, we can say that Lounesto's classification is not unique, and that the six disjoint classes he proposed represent a specific case tied to the use of the Dirac dual. If we consider a deformation of the dual structure, additional classes may become accessible, potentially revealing new types of spinors, provided that the only physical and mathematical requirement namely, the invariance of the bilinear forms under Lorentz transformations and compliance with the FPK identities is preserved, as we can check in \cite{Rogerio:2024lfh}. Within the framework of quantum field theory and mathematical physics, the pursuit of a deeper understanding of the dual structure has significantly motivated the scientific community, especially following the mentioned theoretical discovery of Elko spinors. In light of this, the present work aims to develop a new mathematical approach. Here, we seek a dual structure that incorporates the building blocks of the Clifford algebra and its possible combinations of the identity and $\gamma$ matrices, given by $\Gamma = \lbrace \mathbbm{1}, \gamma_{\mu}, \gamma_{0123}, \gamma_{\mu}\gamma_{0123}, \gamma_{\mu}\gamma_{\nu}\rbrace$. Such basis elements open up the possibility of defining a general structure, a multi-vector in $\mathbb{C} \otimes C \ell_{1,3}$, which composes the dual. This structure carries certain phase factors to be fixed later, either according to the spinor properties or to each class within a general spinorial classification. In doing so, we aim to analyse the invariance of the bilinear forms and determine the phase parameters that must be fixed in each case, both to recover previously established results and to explore new outcomes arising from this mathematical framework.

The paper is organized as follows: In the next section we revisit the bilinear forms taking into account the Dirac dual. In Section \ref{gendual} we set up the framework for a new approach for generalized dual structures, establishing the relationship with previous proposals. In Section \ref{new_hidden} we explore the proposed new dual in the light of the new spinorial classes and find representatives of each of the new class. Finally, in Section \ref{final_remarks} we conclude.
\section{Dirac Dual}
Let $\boldsymbol{\gamma}^{i}$ be Clifford matrices, with $\boldsymbol{\sigma}_{ik}\!=\![\boldsymbol{\gamma}_{i},\boldsymbol{\gamma}_{k}]/4$ the generators of the complex Lorentz group and $2i\boldsymbol{\sigma}_{ab}\!=\!\varepsilon_{abcd}\boldsymbol{\pi}\boldsymbol{\sigma}^{cd}$ implicitly defining the parity-odd matrix $\boldsymbol{\pi}$ (this is usually denoted by a gamma with an index five, but as this is not a true index we prefer to use an index-free notation). The exponentiation of the generators gives the complex Lorentz group $\boldsymbol{\Lambda}$ and then $\boldsymbol{S}\!=\!\boldsymbol{\Lambda}e^{iq\alpha}$ gives the full spinor group. Spinors are objects transforming as $\psi\!\rightarrow\!\boldsymbol{S}\psi$ and $\overline{\psi}\!\rightarrow\!\overline{\psi}\boldsymbol{S}^{-1}$ where $\overline{\psi}\!=\!\psi^{\dagger}\boldsymbol{\gamma}^{0}$ is the adjoint operation. With a pair of adjoint spinors one forms the spinorial bi-linear quantities
\begin{eqnarray}
&\Sigma^{ab}\!=\!2\overline{\psi}\boldsymbol{\sigma}^{ab}\boldsymbol{\pi}\psi\ \ \ \
\ \ \ \ \ \ \ \ M^{ab}\!=\!2i\overline{\psi}\boldsymbol{\sigma}^{ab}\psi\\
&S^{a}\!=\!\overline{\psi}\boldsymbol{\gamma}^{a}\boldsymbol{\pi}\psi\ \ \ \
\ \ \ \ \ \ \ \ U^{a}\!=\!\overline{\psi}\boldsymbol{\gamma}^{a}\psi\\
&\Theta\!=\!i\overline{\psi}\boldsymbol{\pi}\psi\ \ \ \
\ \ \ \ \ \ \ \ \Phi\!=\!\overline{\psi}\psi,
\end{eqnarray}
all of which being real tensors. We have $\Sigma^{ab}\!=\!-\frac{1}{2}\varepsilon^{abij}M_{ij}$ as well as
\begin{eqnarray}
&M_{ab}\Phi\!-\!\Sigma_{ab}\Theta\!=\!U^{j}S^{k}\varepsilon_{jkab}\ \ \ \ \ \ \ \
\ \ \ \ \ \ \ \ M_{ab}\Theta\!+\!\Sigma_{ab}\Phi\!=\!U_{[a}S_{b]}\label{MSigma}
\label{momenta}
\end{eqnarray}
together with
\begin{eqnarray}
&M_{ik}U^{i}=\Theta S_{k}\ \ \ \ \ \ \ \ \Sigma_{ik}U^{i}\!=\!\Phi S_{k}\label{products-1}\\
&M_{ik}S^{i}=\Theta U_{k}\ \ \ \ \ \ \ \ \Sigma_{ik}S^{i}\!=\!\Phi U_{k}\label{products-2}
\end{eqnarray}
and
\begin{eqnarray}
&\frac{1}{2}M_{ab}M^{ab}\!=\!-\frac{1}{2}\Sigma_{ab}\Sigma^{ab}\!=\!\Phi^{2}\!-\!\Theta^{2} \label{orthogonal}\\
&\frac{1}{2}M_{ab}\Sigma^{ab}\!=\!-2\Theta\Phi\label{norm}
\end{eqnarray}
\begin{eqnarray}
&U_{a}U^{a}\!=\!-S_{a}S^{a}\!=\!\Theta^{2}\!+\!\Phi^{2}\label{NORM}\\
&U_{a}S^{a}\!=\!0\label{ORTHOGONAL},
\end{eqnarray}
known Fierz re-arrangements, all being general.

Spinor fields are traditionally classified according to the irreducible representations of the spin group $\text{Spin}^+(1,3)$ connected to the identity, corresponding to the well-known Weyl, Dirac, and Majorana spinors. Alternatively, Lounesto proposed a more algebraically and geometrically appealing classification within the framework of Clifford algebras \cite{Lounesto:2001zz,Vaz:2016qyw}. This classification is based on bilinear covariants and Fierz re-arrangements, as introduced earlier. The classification is summarized in Table \ref{tab_lounesto}, where spinors in the first three classes are referred to as \textit{regular}, while those in the remaining classes are termed \textit{singular}.
\begin{table}[h]
\centering
\begin{tabular*}{.5\linewidth}{@{\extracolsep{\fill}}c|cccccc}
\toprule
\textbf{Lounesto Classes} & $\Phi$ & $\Theta$ & ${U}$ & ${S}$ & ${M}$ \\
\midrule
{1.}   & $\neq0$ & $\neq0$ & $\neq0$ & $\neq0$ & $\neq0$ \\
{2.}  & $\neq0$ & $=0$ & $\neq0$ & $\neq0$ & $\neq0$ \\
{3.}  & $=0$ & $\neq0$ & $\neq0$ & $\neq0$ & $\neq0$ \\
{4.}  & $=0$ & $=0$ & $\neq0$ & $\neq0$ & $\neq0$ \\
{5.}  & $=0$ & $=0$ & $\neq0$ & $=0$ & $\neq0$ \\
{6.}  & $=0$ & $=0$ & $\neq0$ & $\neq0$ & $=0$\\
\bottomrule
\end{tabular*}
\caption{All the standard Lounesto classes.}\label{tab_lounesto}
\end{table}

\noindent Recent investigations have significantly advanced Lounesto's spinor classification, extending its framework to encompass spinors beyond the original six classes. In particular, the reference \cite{Rogerio:2024lfh} shows all the possible classes that serve as extensions of the standard Lounesto classes. The new classes are closely tight to alternative formulations of the Dirac adjoint, as in the seminal work of references \cite{Ahluwalia:2004ab, Ahluwalia:2004sz, Ahluwalia:2016rwl}, which proposes a fundamental approach to describing dark matter \cite{de2025wigner,rogerio2025spinor}. Further formal achievements in Refs. \cite{HoffdaSilva:2019ykt, CoronadoVillalobos:2020yvr, Rogerio:2023kcp, BuenoRogerio:2017lim, HoffdaSilva:2017waf, Rogerio:2022tsl} have also contributed to the development of a general dual for spinors and its classification. The interplay between these extensions will be analysed in Section \ref{new_hidden}, explicitly constructing representatives of spinor classes not covered by Lounesto's original framework. We argue that, beyond the formal advances, these results potentially provide a robust foundation for constructing theories that venture beyond the Standard Model.
\section{General Dual}\label{gendual}
One can define the most general spinor dual by either defining it in general, or by defining it in terms of the Dirac dual according to
\begin{align}
\tilde{\psi}\!=\!\overline{\psi}\Delta,
\end{align}
for some matrix $\Delta$  that has to be a scalar in the tangent space. In Weyl representation, the most general form of $\Delta$ is given by \cite{HoffdaSilva:2019ykt}
\begin{align}
\Delta = \left(\begin{array}{rrrr}
a_{11} & a_{12} & a_{13} & a_{14} \\
a_{21} & a_{22} & \overline{a_{14}} & a_{24} \\
a_{31} & a_{32} & \overline{a_{11}} & \overline{a_{21}} \\
\overline{a_{32}} & a_{42} & \overline{a_{12}} & \overline{a_{22}}
\end{array}\right), \quad \text{with }  a_{13}, a_{31}, a_{24}, a_{42} \in \mathbb{R} \text{ and the overbar denotes de complex conjugation}.
\end{align}

However, this matrix form is not particularly convenient for analyzing how bilinear covariants are influenced by additional parameters. Instead, we aim to preserve the full generality of  $\Delta$ while expressing it in terms of the multivector structure naturally inherited from Clifford Algebras. Regarding the Dirac algebra, it is well known that $\mathbb{C}\otimes C \ell_{1,3}\simeq \mathrm{Mat}_{4\times 4}(\mathbb{C})$. Such isomorphism guarantees that there must exist a multivector in $\mathbb{C}\otimes C \ell_{1,3}$ whose matrix representation matches $\Delta$. We denote this multivector as $\boldsymbol{A}$. Without loss of generality, we can use the time-like vector $v^i$ and the space-like vector $n^i$ that are defined in general in $1 + 1 + 2$ formalisms, so to ensure the scalar structure. With these ingredients, we define

\begin{align}
\boldsymbol{A}\!=\!a\mathbb{I}
\!+\!ib\boldsymbol{\pi}
\!+\!v_{a}\boldsymbol{\gamma}^{a}
\!+\!n_{a}\boldsymbol{\gamma}^{a}\boldsymbol{\pi}
\!+\!i h_{ab}\boldsymbol{\sigma}^{ab},
\end{align}
where $h_{ab}$ is antisymmetric, and the coefficients are to be determined. These coefficients can be constrained by imposing $\boldsymbol{A}$ to have well-defined properties. For instance, requiring that $\boldsymbol{A}$ satisfies the property  $\boldsymbol{\gamma}^{0}\boldsymbol{A}^\dagger\boldsymbol{\gamma}^{0} = \boldsymbol{A}$ \cite{HoffdaSilva:2019ykt} ensures that all coefficients are real. We shall return to the discussion on fixing parameters shortly.

Back to the isomorphism, we observe that $\Delta$ has 16 degrees of freedom, the same as $\boldsymbol{A}$ when its coefficients are taken as real or pure imaginary. In fact, the equivalence between both representations is explicitly realized through the multivector decomposition
\begin{align}
\boldsymbol{A} = \frac 14\underbrace{\mathrm{Tr}(\Delta)\mathbb{I}}_{\rm{scalar}} +\frac 14\underbrace{\mathrm{Tr}(\Delta\boldsymbol{\pi})\boldsymbol{\pi}}_{\rm{pseudoscalar}}+ \frac 14\underbrace{\mathrm{Tr}(\Delta\boldsymbol{\gamma}_a)\boldsymbol{\gamma}^a}_{\rm{vector}} +
 \frac 14\underbrace{\mathrm{Tr}(\Delta\boldsymbol{\gamma}_a\boldsymbol{\pi})\boldsymbol{\pi}\boldsymbol{\gamma}^a}_{\rm{pseudovector}} +\underbrace{\mathrm{Tr}(\Delta\boldsymbol{\sigma}_{ab}^\dagger)\boldsymbol{\sigma}^{ab}}_{\rm{bivector}}.  
\end{align}

Explicit computation of the coefficients in terms of the $\Delta$ parameters yields

{\renewcommand{\arraystretch}{1.5}
\begin{align}
\begin{array}{llllll}  
a & = \frac 14\mathrm{Tr}(\Delta) &= \frac{1}{2} \, \Re (a_{11}) + \frac{1}{2} \, \Re(a_{22}), & 
b & =-\frac{i}{4} \mathrm{Tr}(\Delta\boldsymbol{\pi}) &= \frac{1}{2} \, \Im ( a_{11} ) + \frac{1}{2} \, \Im ( a_{22}),\\
v_{0} & = \frac 14\mathrm{Tr}(\Delta\boldsymbol{\gamma}_0) &=  \frac{1}{4} \, (a_{13} +   a_{24} + a_{31} +  a_{42}), & 
v_{1} & = \frac 14\mathrm{Tr}(\Delta\boldsymbol{\gamma}_1) &= \frac{1}{2} \, \Re ( a_{32} ) -\frac{1}{2} \, \Re ( a_{14} ), \\
v_{2} & = \frac 14\mathrm{Tr}(\Delta\boldsymbol{\gamma}_2) &=  \frac{1}{2} \, \Im ( a_{14} ) - \frac{1}{2} \, \Im ( a_{32} ), & 
v_{3} & = \frac 14\mathrm{Tr}(\Delta\boldsymbol{\gamma}_3) &=  -\frac{1}{4} \,( a_{13} - a_{24} - a_{31} +  a_{42}),\\
n_{0} & = \frac 14\mathrm{Tr}(\Delta\boldsymbol{\gamma}_0\boldsymbol{\pi}) &=  \frac{1}{4} \, (a_{13} +  a_{24} -  a_{31} -  a_{42}),& 
n_{1} & = \frac 14\mathrm{Tr}(\Delta\boldsymbol{\gamma}_1\boldsymbol{\pi}) &=  -\frac{1}{2} \, \Re ( a_{14} ) - \frac{1}{2} \, \Re ( a_{32} ),\\
n_{2} & = \frac 14\mathrm{Tr}(\Delta\boldsymbol{\gamma}_2\boldsymbol{\pi}) &=  \frac{1}{2} \, \Im( a_{14} ) + \frac{1}{2} \, \Im ( a_{32} ),& 
n_{3} & = \frac 14\mathrm{Tr}(\Delta\boldsymbol{\gamma}_3\boldsymbol{\pi}) &=  -\frac{1}{4} \,( a_{13} - a_{24} + a_{31} - a_{42}),\\
h_{01} & = -i\mathrm{Tr}(\Delta\boldsymbol{\sigma}_{01}^\dagger) &=  \Im ( a_{12} ) + \Im ( a_{21} ), & 
h_{02} & = -i\mathrm{Tr}(\Delta\boldsymbol{\sigma}_{02}^\dagger) &=  \Re ( a_{12} ) - \Re ( a_{21} ),\\
h_{03} & = -i\mathrm{Tr}(\Delta\boldsymbol{\sigma}_{03}^\dagger) &=  \Im ( a_{11} ) - \Im ( a_{22} ), &
h_{12} & = -i\mathrm{Tr}(\Delta\boldsymbol{\sigma}_{12}^\dagger) &=  \Re ( a_{11} ) - \Re ( a_{22} ),\\
h_{23} & = -i\mathrm{Tr}(\Delta\boldsymbol{\sigma}_{23}^\dagger) &=  \Re ( a_{12} ) + \Re ( a_{21} ), &
h_{31} & = -i\mathrm{Tr}(\Delta\boldsymbol{\sigma}_{31}^\dagger) &=  \Im ( a_{21} ) -\Im ( a_{12} ). 
\end{array}
\end{align}}
which confirms that all coefficients of $\boldsymbol{A}$ are real. Thus $\boldsymbol{A}$ retains the full generality of      $\Delta$, while offering a manifestly covariant and computationally convenient representation to handle with bilinear covariants. For completeness, we can also express $\Delta$ in terms of $\boldsymbol{A}$ coefficients, as given in the matrix
\begin{align}
\!\!\!\!\Delta\!=\!\begin{pmatrix}
a\!+\!ib\!+\!\frac 12(h_{12}\!+\!ih_{03})&\frac 12[h_{02}\!-\!h_{23} \!+\! i(h_{01}\!-\!h_{31})]&v_0\!-\!v_3\!+\!n_0\!-\!n_3&\!-\!v_{1}\!-\!n_{1} \!+\! i(v_{2}\!+\!n_{2})\\
\frac 12[h_{23}\!-\!h_{02} \!+\! i(h_{01}\!+\!h_{31})]&a\!+\!ib\!-\!\frac 12(h_{12}\!+\!ih_{03})&\!-\!v_{1}\!-\!n_{1} \!-\! i(v_{2}\!+\!n_{2})&v_0\!+\!v_3\!+\!n_0\!+\!n_3\\
v_0\!+\!v_3\!-\!n_0\!-\!n_3&v_{1}\!-\!n_{1} \!-\! i(v_{2}\!-\!n_{2})&a\!-\!ib\!+\!\frac 12(h_{12}\!-\!ih_{03})& \frac 12[h_{23}\!-\!h_{02} \!-\! i(h_{01}\!+\!h_{31})]\\
v_{1}\!-\!n_{1} \!+\! i(v_{2}\!-\!n_{2})&v_0\!-\!v_3\!-\!n_0\!+\!n_3&\frac 12[h_{02}\!-\!h_{23} \!-\! i(h_{01}\!-\!h_{31})]&a\!-\!ib\!-\!\frac 12(h_{12}\!-\!ih_{03})
\end{pmatrix}.
\end{align}
Now, using $\boldsymbol{A}$ and the most general dual given by $\tilde{\psi} = \bar{\psi}A$, we are able to find the new general bilinears in terms of the Dirac bilinears
\begin{align}
\tilde{M}^{j k} &= a M^{j k} -b \Sigma^{j k} + i v^{[j}U^{k]} +  v_a \epsilon^{a j k b}S_b  +
 i n^{[j}S^{k]} + n_a \epsilon^{a j k b} U_b + i  M^{a [j}h^{k]}{_{a}}+ h^{j k} \Phi -\frac 12 h_{a b}\epsilon^{a b j k} \Theta;\\
 \tilde{S}^j &= a S^j - i b U^j -i v^j\Theta + v_a\Sigma^{a j} - n^j\Phi + i n_a M^{a j} + 
 i h^{a j}S_a + \frac 12 h_{a b}\epsilon^{a b j k}U_k;\\
 \tilde{U}^j &=  a U^j - i b S^j +  v^{j} \Phi + i v_aM^{j a} + i n^{j} \Theta +  n_a\Sigma^{j a} +  
 i h^{a j} U_a + \frac 12 h_{a b} \epsilon^{a b j k} S_k;\\
 \tilde{\Theta} &= a \Theta - b \Phi + i v_a S^a + i n_a U^a - \frac{1}{2}h_{a b} \Sigma^{a b};\\
\tilde{\Phi} &=  a \Phi + b \Theta + v_a U^a + n_a S^a + \frac{1}{2} h_{a b} M^{a b};
\end{align}
as a straightforward computation would show. If they are all required to be real, we would obtain that all coefficients
would vanish except $a$, which would have to be real, and so we would reduce to the Dirac case (up to one normalization
constant): hence, if we want to enlarge our description, some of these bi-linears should be allowed to be complex. On
the other hand, in this case, they must be purely imaginary, so that they can be reduced to be real after multiplying
by the imaginary unit. So, for example, in the complementary case $a = b = 0$ and $h_{ab} = 0$, we would have that 
$\tilde{\Phi} = v_a U^a + n_a S^a$ would be real and $\tilde{\Theta} = i v_a S^a + i n_a U^a $ would be complex, although 
$-i\tilde{\Theta}$ would be real again. Notice, however, that $\tilde{M}^{j k}, \tilde{S}^j$ and $\tilde{U}^j$ would still be constituted by both real and imaginary parts. So that there is, in general, no manner with which to reduce all bilinears to be real, unless some additional constraint is taken. In the following, we shall analyze some notable constraints in some specific cases of spinor fields.

We now introduce the parameters $c,d,e$ and express $\boldsymbol{A}$ as
\begin{align}
\boldsymbol{A}\!=\!a\mathbb{I}
\!+\!ib\boldsymbol{\pi}
\!+\!cv_{a}\boldsymbol{\gamma}^{a}
\!+\!dn_{a}\boldsymbol{\gamma}^{a}\boldsymbol{\pi}
\!+\!ie h_{ab}\boldsymbol{\sigma}^{ab}.
\end{align}
Notice that these parameters does not introduce new degrees of freedom, only rescale the previous one. However, they will be useful in what follows. Considering the constraint $\boldsymbol{A}^2 = \mathbb{I}$, we find
\begin{align}
    a^2 - b^2 + c^2 + d^2 - e^2 &= 1, \\
    ab &= 0, \label{eq:constraint2} \\
    ac &= 0, \label{eq:constraint3} \\
    ad &= 0, \label{eq:constraint4} \\
    ae &= 0, \label{eq:constraint5} \\
    cd - be &= 0. \label{eq:constraint6}
\end{align}
which leaves two choices: $b = c = d = e = 0$ and $a$ unitary (i.e., the Dirac case), or $a = 0$ with  
\begin{align}
    c &= \cosh \alpha - d,  \\
    b &= \sinh \alpha - e, 
\end{align}
and the additional constraint $(\cosh \alpha - d)d = (\sinh \alpha - e)e.$ One solution is, for example, $b = d = 0$ with $c = \cosh \alpha$ and $e = \sinh \alpha$, then $\boldsymbol{A} = \cosh \alpha v_a \boldsymbol{\gamma}^a + i \sinh \alpha v_{[a} n_{b]} \boldsymbol{\sigma}^{ab}$. We will explore this case below. Let us now consider the case of Majorana spinors, subject to the constraints \( \Phi = \Theta = 0 \) and \( S_a = 0 \): in this circumstance,  
some bilinears are still not real. However, we may now make two fully complementary cases: the first is given by  
\( c = e = 0 \), for which we obtain that \( \tilde{M}^{ij} \), \( i\tilde{S}^i \), \( \tilde{U}^j \), and \( -i\tilde{\Theta} \) would all be real, but \( \tilde{\Phi} = 0 \) identically.  
More interesting is the case \( a = b = d = 0 \), for which  

\begin{align}
-i\tilde{M}^{j k} &=   cv^{[j}U^{k]}   +  eM^{a [j}h^{k]}{_{a}};\\
 \tilde{S}^j &=   \frac{c}{2}v_a\varepsilon^{jabk}M_{bk} + \frac{e}{2} h_{a b}\epsilon^{a b j k}U_k;\\
 i\tilde{U}^j &=    cv_aM^{a j}  - eh^{a j} U_a \\
 \tilde{\Theta} &=    \frac{e}{4}h_{a b} \varepsilon^{abjk}M_{jk};\\
\tilde{\Phi} &=   cv_a U^a  + \frac{e}{2} h_{a b} M^{a b};
\end{align}
all being real, and none zero, in general. Notice that this case falls precisely into the example \( c = \cosh \alpha \) and \( e = \sinh \alpha \)  
seen above. We regard this as the most general way in which the dual can be defined for Majorana spinors, while  
maintaining manifest covariance and reality for all the bilinears. 

Another simple set of bilinears can be achieved by taking
\begin{align}
 \boldsymbol{A} = a \mathbb{I} + ib\boldsymbol{\pi} + (c_1 U_a+c_2 S_a)\boldsymbol{\gamma}^a+(d_1 U_a+d_2 S_a)\boldsymbol{\gamma}^a\boldsymbol{\pi} + ieU_{[a}S_{b]}\boldsymbol{\sigma}^{a b}.
\end{align}
After defining $p = c_1-e\Theta-d_2$ and $q = d_1-ie\Phi-c_2$, the bilinear can be expressed as
\begin{align}\label{eq:new_regulars_0}
 \tilde{M}^{j k} &= a {M}^{j k} -b\Sigma^{j k}-i q U^{[k}S^{j]},\\
 \tilde{S}^j &= (a+p\Phi+iq\Theta)S^j -i(b-iq\Phi+p\Theta)U^j,\\
 \tilde{U}^j &= (a+p\Phi+iq\Theta)U^j -i(b-iq\Phi+p\Theta)S^j, \\
 \tilde{\Theta} &= a \Theta -b\Phi  + iq(\Phi^2+\Theta^2),\\ \label{eq:new_regularsf_0}
\tilde{\Phi} &= a \Phi +b \Theta + p(\Phi^2+\Theta^2).
\end{align}
We find a set of real regular bilinears by choosing $iq=1$ and $b-iq\Phi+p\Theta=0$. Thus
\begin{align}\label{eq:new_regulars_1}
 \tilde{M}^{j k} &= a {M}^{j k} -b\Sigma^{j k}- U^{[k}S^{j]},\\ \label{eq:new_regularsf_1}
 \tilde{S}^j &= \omega S^j,\quad  \tilde{U}^j = \omega U^j, \quad \tilde{\Theta} = \omega \Theta,\quad \tilde{\Phi} = \omega \Phi.
\end{align}
For $\omega = \left[ a+\frac{\Phi}{\Theta}(\Phi-b)+\Theta\right].$ It is straightforward to see that, taking $\omega=0$, the resulting spinor does not fit in any of the standard Lounesto classes displayed in Table \ref{tab_lounesto}. We will return to this discussion in Section \ref{new_hidden}.

It is important to emphasize that, depending on the choice of the free parameters in $\textbf{A}$, we are able to recover the results obtained in \cite{Rogerio:2024lfh}. In other words, the operator $\textbf{A}$ can play the role of identity, parity, charge conjugation, or any other discrete symmetry. In this way, we can understand $\textbf{A}$ as a more general operator, whose particular cases correspond to the discrete symmetries.
\section{New bilinears and hidden classes}\label{new_hidden}
In this section we explore the connection between the general bilinears as discussed in previous sections and the general extension of Lounesto's classification introduced in \cite{Rogerio:2024lfh}. While Lounesto's original framework was primarily based on the Dirac spinor, a more comprehensive classification was found as result of a careful examination of the constraints imposed by the FPK identities, which dictate the permissible sets of bilinear covariants and, consequently, the allowed spinor classes. After reviewing these constraints, the resulting set of admissible subclasses can potentially serve as the foundation for any spinor theory consistent with the FPK identities. The general extension of Lounesto's obeying FPK identities is shown in Table \ref{tab_allowed} below. Any possibility not included here is forbidden by the mentioned identities.

\begin{table}[h]
\centering
\begin{tabular*}{.5\linewidth}{@{\extracolsep{\fill}}c|cccccc}
\toprule
\textbf{Allowed classes} & $\Phi$ & $\Theta$ & ${U}$ & ${S}$ & ${M}$\\
\midrule
\textbf{1.}   & $\neq0$ & $\neq0$ & $\neq0$ & $\neq0$ & $\neq0$ \\
\emph{1.1}   & $\neq0$ & $\neq0$ & $\neq0$ & $\neq0$ & $=0$ \\
\emph{1.2}   & $\neq0$ & $\neq0$ & $\neq0$ & $=0$ & $\neq0$ \\
\emph{1.3}   & $\neq0$ & $\neq0$ & $\neq0$ & $=0$ & $=0$ \\
\emph{1.4}   & $\neq0$ & $\neq0$ & $=0$ & $\neq0$ & $\neq0$ \\
\emph{1.5}   & $\neq0$ & $\neq0$ & $=0$ & $=0$ & $=0$ \\
\emph{1.6}   & $\neq0$ & $\neq0$ & $=0$ & $=0$ & $\neq0$ \\
\emph{1.7}   & $\neq0$ & $\neq0$ & $=0$ & $\neq0$ & $=0$ \\
\textbf{2.}  & $\neq0$ & $=0$ & $\neq0$ & $\neq0$ & $\neq0$ \\
\emph{2.1}   & $\neq0$ & $=0$ & $\neq0$ & $\neq0$ & $=0$ \\
\textbf{3.}  & $=0$ & $\neq0$ & $\neq0$ & $\neq0$ & $\neq0$ \\
\emph{3.1}   & $=0$ & $\neq0$ & $\neq0$ & $\neq0$ & $=0$ \\
\textbf{4.}  & $=0$ & $=0$ & $\neq0$ & $\neq0$ & $\neq0$ \\
\emph{4.1}   & $=0$ & $=0$ & $=0$ & $\neq0$ & $\neq0$ \\
\textbf{5.}  & $=0$ & $=0$ & $\neq0$ & $=0$ & $\neq0$ \\
\emph{5.1}   & $=0$ & $=0$ & $=0$ & $=0$ & $\neq0$ \\
\textbf{6.}  & $=0$ & $=0$ & $\neq0$ & $\neq0$ & $=0$ \\
\emph{6.1}   & $=0$ & $=0$ & $=0$ & $\neq0$ & $=0$ \\
\textbf{7.}  & $=0$ & $=0$ & $\neq0$ & $=0$ & $=0$ \\
\bottomrule
\end{tabular*}
\caption{All the classes allowed according to FPK identities, taken from \cite{Rogerio:2024lfh}.}\label{tab_allowed}
\end{table}

Despite being allowed by the FPK identities, there was no explicit proof that the new classes are not empty. Identifying explicit representatives of the new classes is not straightforward due to the intricate interrelations among bilinear covariants. Our aim in this Section is to pinpoint such representatives by making appropriate choices for the coefficients of $\boldsymbol{A}$. The standard classes can always be accessed by taking $a=1$ and all the other coefficients being null, i.e. falling into Dirac's bilinears. It is worth remembering that, given a set of bilinears, the corresponding Dirac spinor can be recovered by applying the Takahashi's inversion theorem \cite{Takahashi:1982bb}. We begin by the singular classes extensions.

\subsection{Singular Extensions}

For the first two singular classes we take the bilinears introduces in equations \eqref{eq:new_regulars_0}$-$\eqref{eq:new_regularsf_0}. Considering the most general singular seed spinor in standard Lounesto classification, namely, the one of Class 4, it yields
\begin{align}\label{eq:new_singulars}
\tilde{M}^{j k} &= a {M}^{j k} -b\Sigma^{j k}-i q U^{[k}S^{j]},\\
\tilde{S}^j &= a S^j -ib U^j,\\
\tilde{U}^j &= a U^j -ib S^j, \\
\tilde{\Theta} &= 0,\\ \label{eq:new_singularsf}
\tilde{\Phi} &= 0.
\end{align}

Depending on the choice of parameters and Dirac bilinears class, it allows us to construct representatives for the singular extensions as follows:

\begin{itemize}

\item[\textbf{Class 4.1:}]
A representative of this class can be obtained by setting $a = 0$ and using the Dirac bilinear of Class 5. The non-vanishing new bilinears simplify to:
\begin{align}
\tilde{M}^{jk} &= -b \Sigma^{jk}, \\
i \tilde{S}^j &= b U^j,
\end{align}
while all others vanish. This satisfies the conditions of Class 4.1.
Alternatively, the same representative is recovered by choosing a nonzero $b$ and setting $v_a = U_a$ as the only nontrivial components in $\boldsymbol{A}$. Again starting with a Class 5 spinor.

\item[\textbf{Class 5.1:}]
To find a straightforward representative of Class 5.1, we take $\omega=0$ in the set \eqref{eq:new_regulars_1}$-$\eqref{eq:new_regularsf_1}. Which is sufficient to characterize this Class.

\item[\textbf{Class 6.1:}]
We now turn to Dirac bilinears in Class 6 and the new bilinears given by equations \eqref{eq:new_regulars_0}$-$\eqref{eq:new_regularsf_0}. It yields
\begin{align}\label{eq:new_singulars2}
\tilde{M}^{jk} &= 0, \\
\tilde{S}^j &= a S^j - i b U^j, \\
\tilde{U}^j &= a U^j - i b S^j, \\
\tilde{\Theta} &= 0, \\ \label{eq:new_singularsf2}
\tilde{\Phi} &= 0.
\end{align}
By imposing the condition $S^j = -i \frac{a}{b} U^j$ we obtain a valid representative of Class 6.1. In this case, we simply redefine  $S$ as $S\to iS$ in order to keep it real. It is worth noting that this choice preserves the Fierz re-arrangement identities.

\item[\textbf{Class 7:}]
Using the same set of bilinears as above, but now imposing the constraint $S^j = i \frac{b}{a} U^j$ we construct a spinor of Class 7. This configuration is dual to the previous one and completes the singular classification under this framework.

\end{itemize}

\subsection{Regular Extensions}

For extensions of the first regular classes we shall use a simple generalisation of the Dirac dual, given by $\boldsymbol{A}=a\mathbb{I} +ib\boldsymbol{\pi}$. In this case, the new bilinears are simply:
\begin{align}
\tilde{M}^{j k} &= a M^{j k} -b \Sigma^{j k};\\
 \tilde{S}^j &= a S^j - i b U^j;\\
 \tilde{U}^j &=  a U^j - i b S^j ;\\
 \tilde{\Theta} &= a \Theta - b \Phi ;\\
\tilde{\Phi} &=  a \Phi + b \Theta.
\end{align}
Explicit representatives for each subclass follow from the above bilinears and are given below:
\begin{itemize}
 \item[\textbf{Class 1.1:}] For this class, $\tilde{M}^{jk}=0$ implies that $M^{jk}$ is proportional to its dual, i.e., $M^{jk}= - \frac{b}{2a}\epsilon^{j k c d}M_{cd}$. While all others are non null.

 \item[\textbf{Class 1.2:}] Enforcing $\tilde{S}^j = 0$ yields the constraint $S^j = i\frac{b}{a}U^j$. As previously, we redefine  $S$ as $S\to iS$ in order to keep it real.

  \item[\textbf{Class 1.3:}] This subclass requires both previous conditions: $M^{jk}= - \frac{b}{2a}\epsilon^{j k c d}M_{cd}$ and $S^j = i\frac{b}{a}U^j$.

 \item[\textbf{Class 1.4:}] Analogous to Class 1.2, with the roles of $S^j$ and $U^j$ exchanged: $U^j = i\frac{b}{a} S^j.$

  \item[\textbf{Class 1.5:}] This is the most restrictive class, yet a representative can be constructed by combining the constraints from Classes 1.1 and 1.2, as well as relaxing the reality condition on the coefficients of $\boldsymbol{A}$. Specifically, by taking $a=ib$.

  \item[\textbf{Class 1.6:}] For class $1.6$ we consider the Dirac bilinears belonging to class 6 and fix the coefficients as $a = 0$, $b = 0$ and $h_{ab} = 0$. Under these assumptions, we have
\begin{align*}
\tilde{\Phi} &= v_a U^a + n_a S^a, \\
\tilde{\Theta} &= i n_a U^a + i v_a S^a, \\
\tilde{U}^j &= 0, \\
\tilde{S}^j &= 0, \\
\tilde{M}^{jk} &= i v^{[j} U^{k]} + i n^{[j} S^{k]}+ n_a \varepsilon^{ajkb} U_b + v_a \varepsilon^{ajkb} S_b.
\end{align*}

Which satisfy the defining conditions of class $1.6$.

   \item[\textbf{Class 1.7:}] This subclass combines the conditions of Classes 1.1 and 1.4: $M^{jk}= - \frac{b}{2a}\epsilon^{j k c d}M_{cd}$ and $U^j = i\frac{b}{a}S^j$.
\end{itemize}

  For the remaining regular classes, we consider the extended bilinears defined by equations \eqref{eq:new_regulars_0}$-$\eqref{eq:new_regularsf_0}, under the assumption $c_1 = c_2 = d_1 = d_2 = 0$. The resulting bilinears take the form:
\begin{align}
\tilde{M}^{jk} &= a M^{jk} - b \Sigma^{jk} - e \Phi\, U^{[k} S^{j]}, \\
\tilde{S}^j &= a\, S^j - i\left[b - e(\Phi^2 + \Theta^2)\right] U^j, \\
\tilde{U}^j &= a\, U^j - i\left[b - e(\Phi^2 + \Theta^2)\right] S^j, \\
\tilde{\Theta} &= a \Theta - b \Phi + e \Phi (\Phi^2 + \Theta^2), \\
\tilde{\Phi} &= a \Phi + b \Theta - e \Theta (\Phi^2 + \Theta^2).
\end{align}
The corresponding representatives for the extended classes are:

\begin{itemize}
\item[\textbf{Class 2.1:}] For this class we depart from Dirac bilinears of class 2 and take $e=\frac{b}{\Phi^2}$. It ensures that $\tilde{\Theta}=0$. Additionally, using Fierz identities, one finds:
\begin{align}
\tilde{M}^{jk} = a M^{jk} - b \Sigma^{jk} - e \Phi\, U^{[k} S^{j]} = a M^{jk} - 2b \Sigma^{jk},
\end{align}
leading to the constraint
\begin{align}
M^{jk} = -\frac{b}{a} \epsilon^{jkcd} M_{cd}.
\end{align}

\item[\textbf{Class 3.1:}] Analogously, we depart from Dirac bilinears of class 3,  set $e = \frac{b}{\Theta^2}$ to ensure $\tilde{\Phi} = 0$ and impose

\begin{align}
M^{jk} = -\frac{b}{2a} \epsilon^{jkcd} M_{cd}.
\end{align}
 This completes the regular classification under this framework.
\end{itemize}
\section{Final remarks}\label{final_remarks}
As we have seen throughout this work, it is possible to redefine the dual structure in a mathematically concise covariant manner while preserving the physical information carried by the theory. The need for a well-defined dual has been the subject of recent studies, which show that such a structure must indeed be carefully analysed to ensure that issues arising during quantization such as energy level measurements, zero-point energy, locality, and particle states, for example are always consistent with the postulates of quantum mechanics and quantum field theory. In this way, the approach developed in this work allows us to recover already known results, as well as to interpret recent findings in a highly practical manner, thereby demonstrating the utility of the mechanism we have developed here.
It is also worth emphasizing that the structure we introduced in the dual, depending on the choice of the free parameters, can reproduce, for example, the usual Dirac case, where the structure $\textbf{A}$ is the identity matrix. However, $\textbf{A}$ can also correspond to any other operator, such as the discrete symmetry operators, which were the subject of study in \cite{Rogerio:2024lfh,Rogerio:2023kcp}. More interestingly, for any given set of non-standard bilinears$-$potentially arising from theories beyond the Standard Model$-$the approach introduced here allows for the identification of the associated dual and the reconstruction of the original spinor, with the help of the classical Takahashi's result \cite{Takahashi:1982bb}. This reveals new possibilities that remain concealed within the framework of the standard Dirac dual.
\vspace{10pt}

\textbf{Funding}. RJBR thanks the generous hospitality offered by UNIFAAT and to Prof. Renato Medina. RTC thanks the National Council for Scientific and Technological Development -- CNPq (Grant  No. 401567/2023-0 [Edital Universal]). LF thanks the Next Generation EU through via the project ``Geometrical and Topological effects on Quantum Matter (GeTOnQuaM)''.

\

\textbf{Data availability}. The manuscript will have associated data in any repository.

\

\textbf{Conflict of interest}. There is no conflict of interest.

\end{document}